\documentclass[article,aps,twocolumn,showpacs,amsmath,amssymb,superscriptaddress,nofootinbib,floatfix]{revtex4-1}

\usepackage{graphicx}% Include figure files
\usepackage{dcolumn}% Align table columns on decimal point
\usepackage{bm}% bold math
\usepackage[colorlinks=true,allcolors=blue]{hyperref}% add hypertext capabilities
\usepackage{array,booktabs,tabularx}
\usepackage[caption=false]{subfig}
\usepackage[USenglish]{babel}
\usepackage{braket}
\usepackage{xcolor}
\usepackage{amsfonts}
\usepackage{amssymb}
\usepackage[normalem]{ulem}
\usepackage{notes2bib}

\begin{document}

\title{Order-disorder charge density wave instability in the kagome metal (Cs,Rb)V$_3$Sb$_5$}
\author{D. Subires}
\affiliation{Donostia International Physics Center (DIPC), San Sebastián, Spain}
\author{A. Korshunov}
\affiliation{European Synchrotron Radiation Facility (ESRF), BP 220, F-38043 Grenoble Cedex, France}
\author{A. H. Said}
\affiliation{Advanced Photon Source, Argonne National Laboratory, Lemont, IL 60439}
\author{L. Sánchez}
\affiliation{Donostia International Physics Center (DIPC), San Sebastián, Spain}
\author{Brenden R. Ortiz}
\affiliation{Materials Department and California Nanosystems Institute, university of California Santa Barbara, Santa Barbara, California, 93106, USA}
\author{Stephen D. Wilson}
\affiliation{Materials Department and California Nanosystems Institute, university of California Santa Barbara, Santa Barbara, California, 93106, USA}
\author{A. Bosak}
\affiliation{European Synchrotron Radiation Facility (ESRF), BP 220, F-38043 Grenoble Cedex, France}
\author{S. Blanco-Canosa}
\email{sblanco@dipc.org}
\affiliation{Donostia International Physics Center (DIPC), San Sebastián, Spain}
\affiliation{IKERBASQUE, Basque Foundation for Science, 48013 Bilbao, Spain}

\date{November 2022}

\begin{abstract}
\section{Abstract}

The origin of the charge density wave phases in the kagome metal compound AV$_3$Sb$_5$ is still under great scrutiny. Here, we combine diffuse and inelastic x-ray scattering to identify a 3-dimensional precursor of the charge order at the \textit{L} point that condenses into a CDW through a first order phase transition. The quasi-elastic critical scattering indicates that the dominant contribution to the diffuse precursor is the elastic central peak without phonon softening. However, the inelastic spectra show a small broadening of the Einstein-type phonon mode on approaching T$_{CDW}$. Our results point to the situation where the Fermi surface instability at the \textit{L} point is of order-disorder type with critical growth of quasi-static domains. The experimental data indicate that the CDW consists on an alternating Star of David and trihexagonal distortions and its dynamics goes beyond the classical weak-coupling scenario and is discussed within strong-electron phonon coupling and non-adiabatic models.  
\end{abstract}

\maketitle
\section{Introduction}
The intricate relationship between lattice geometry and topological electronic behavior determines the ground state properties of materials. The non-trivial band topology of the kagome lattice is being extensively explored as candidates to engineer Dirac fermions \cite{Ye18}, topological flat bands \cite{Kan20}, magnetic Weyl semimetals \cite{Liu18} or quantum spin liquid behavior \cite{Yan11,Han12}. Besides, the combination between the non-trivial topology and strong electronic correlations, originated from the interplay of orbital, charge and spin degrees of freedom, introduces topological correlated physics \cite{Mazin2014} with more exotic phenomena, like topological Majorana modes. 

Recently, a new family of non magnetic kagome metals AV$_3$Sb$_5$ (A=K, Rb, Cs) has emerged as a fertile playground to investigate correlation-driven topological phases \cite{Ort19,Ort20,Ort21}. Derived from the kagome structure, this family of materials features a quasi 2D electronic band structure with van Hove singularities (vHs) and Dirac crossings close to the Fermi level \cite{Kan22,Lou22}. Superconductivity sets in at low temperature and is intertwined with an exotic charge density wave phase (CDW) \cite{Zhao2021}. Although the electronic structure obtained from density functional theory (DFT) calculations is well established, controversy surrounds the origin and stabilization of the CDW, and the mechanism of superconductivity and its gap symmetry \cite{Zhao21,Tan21,Wu21}.

The rich phase diagram of the kagome lattice was investigated with respect to Fermi surface instabilities within the Hubbard model, revealing spin and CDW orders and unconventional superconductivity emerging upon tuning the band filling to the vHs \cite{Kie13,Wang13}. Diffraction experiments and scanning tunneling microscopy (STM) unraveled multiple-\textit{q} CDW modulations with an inverse Star of David structure (SoD) and a trihexagonal (TrH) distortions \cite{Miao22,Jian21}, following the phonon instabilities at the \textit{M} and \textit{L} points predicted by DFT \cite{Tan21,Subedi22}. The \textit{q} vector at \textit{M} connects neighbouring vHs in the vicinity of the Fermi surface of AV$_3$Sb$_5$, suggesting that the CDW is triggered by the Peierls mechanism \cite{Kan22,Lou22}.  
On the other hand, the CDW of CsV$_3$Sb$_5$ has been shown to break the time reversal symmetry \cite{Mie22}, unveiling an entanglement of superconductivity and charge order chirality \cite{Jian21}. Nevertheless, the chiral CDW has been questioned by recent STM experiments \cite{Li22}, pointing to a sample/surface dependent fermiology or the presence of dynamical disorder. This is particularly appealing, since the sample quality and A-site occupancy has a strong impact on the T$_c$ \cite{Ort21} and might also influence the possible origin and stabilization of the charge modulations. Further, recent experiments reported a new unidirectional modulation charge stripe order with 4\textit{a}$_0$ wavelength that breaks the rotational symmetry \cite{Li22}, nematicity \cite{Nie22} and pair density wave order \cite{Chen21}.

Central to the interplay between CDW and $\mathbb{Z}_2$ band topology is the nature of the CDW phase. From the thermodynamic point of view, specific heat measurements \cite{Ort20,Ort21} show a small release of entropy through the CDW transition, suggesting a first order phase transition, and tight binding models describe the CDW as electronically mediated \cite{Den21,Fen21}. In contrast, DFT predicts soft modes at \textit{M} and \textit{L} points of the Brillouin zone \cite{Tan21}, for which, however, inelastic x-ray scattering (IXS) does not observe any anomaly of the low energy acoustic branches \cite{Li22Miao}. However, ARPES data have reported deviations from the linear electronic dispersion \cite{Luo22}, providing that the electron-phonon coupling gives sufficient strength for the CDW formation. Furthermore, optical spectroscopy ascribed the opening of the CDW gap to the suppression of density of states at the \textit{M} point \cite{Zhou21} with strong phonon renormalizations \cite{Uyk22} without electronic anomalies at \textit{L}. Finally, in addition to the enigmatic driving force of the CDW, the multiple \textit{q}$_{CDW}$ and the different energy scales \cite{Nak21} hinder a detailed study of the ground state of AV$_3$Sb$_5$, and the origin and stabilization of the CDW phase transition remains unclear.

Here, we use a combination of diffuse scattering (DS) and inelastic x-ray scattering (IXS) to identify frozen phonon fluctuations of the 2$\times$2$\times$2 (\textit{L} point) 3D-CDW in AV$_3$Sb$_5$ at T$>$T$_{CDW}$ that condense into a triple-\textit{Q}$_L$ CDW through a first order phase transition, preserving the six-fold symmetry of the kagome lattice. The precursor of the charge order is absent for the 2$\times$2 (\textit{M} point) and 2$\times$2$\times$4 superstructures, indicating that the 2$\times$2$\times$2 charge modulation is the leading Fermi surface instability. Furthermore, the intensity of the quasi-elastic central peak (CP) of IXS follows the diffuse scattering (DS) intensity, showing that the CDW is of an order-disorder transformation type. This is corroborated by the absence of a clear softening, but small broadening of the optical branch at \textit{L}. Our results solve the dispute about the primary order parameter and highlight the critical role of the order-disorder phase transition to understand the controversial experimental reports.  

\begin{figure}
\begin{center}
\includegraphics[width=\columnwidth,draft=false]{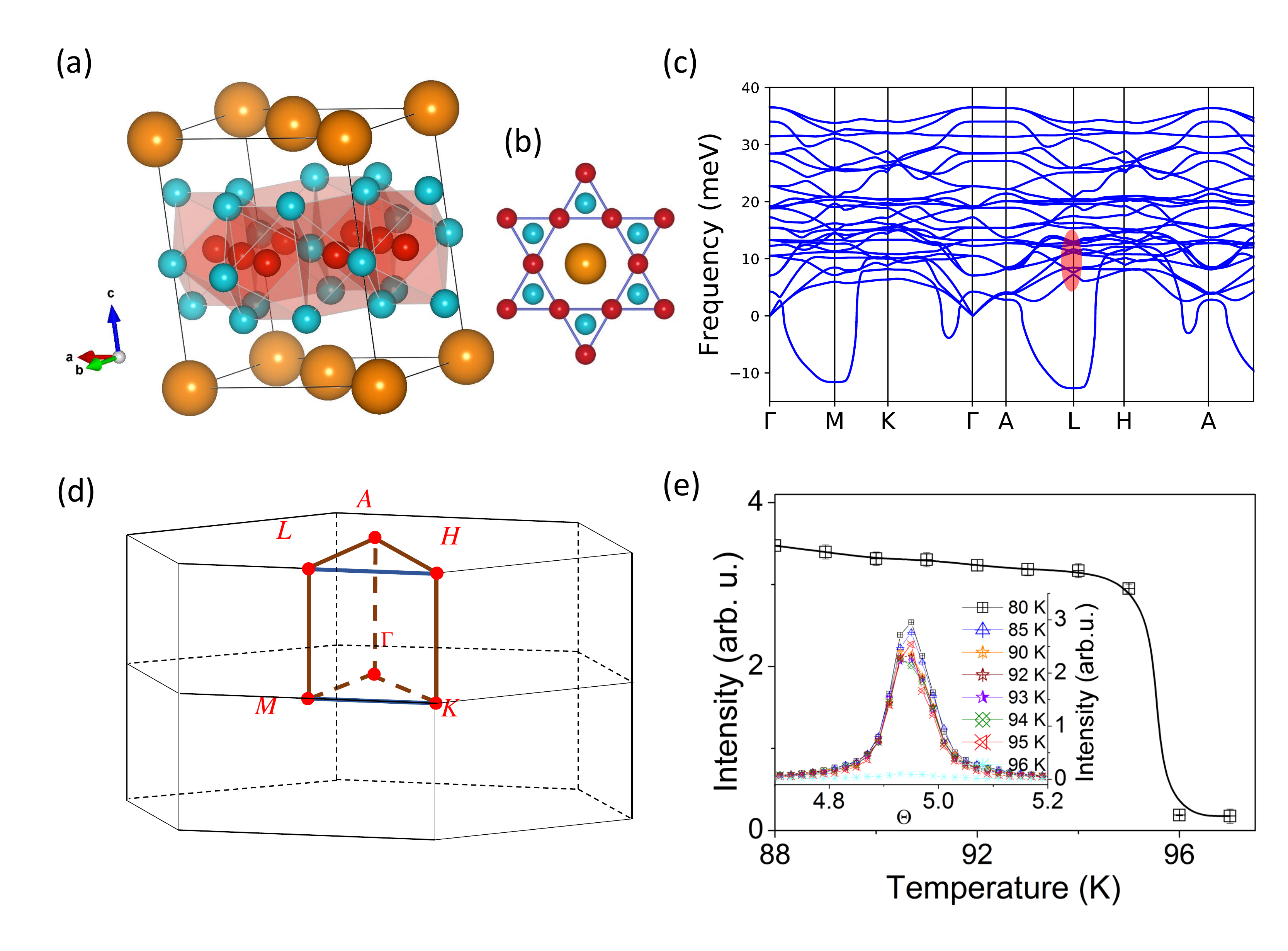}
\caption{\textbf{Charge density wave in CsV$_3$Sb$_5$}. (a) Chemical structure of AV$_3$Sb$_5$. Orange, red and cyan balls stand for A (Cs, Rb), V and Sb atoms. (b) Top view of the kagome structure, showing the V-kagome net interlaced with the Sb atoms. (c) Calculated phonon dispersions of fully-relaxed CsV$_3$Sb$_5$. Soft modes are observed at \textit{M} and \textit{L}. The red ellipse marks the region where phonons are measured in figure 4(d). (d) Sketch of the Brillouin zone along the main symmetry directions. (e) Temperature dependence of the (2.5 0.5 0.5) CDW reflection in CsV$_3$Sb$_5$, showing a sharp drop of intensity at the T$_{CDW}$=95 K. (Inset, rocking scans).}
\label{Fig1}
\end{center}
\end{figure}

\begin{figure*}
\begin{center}
\includegraphics[width=2.0\columnwidth,draft=false]{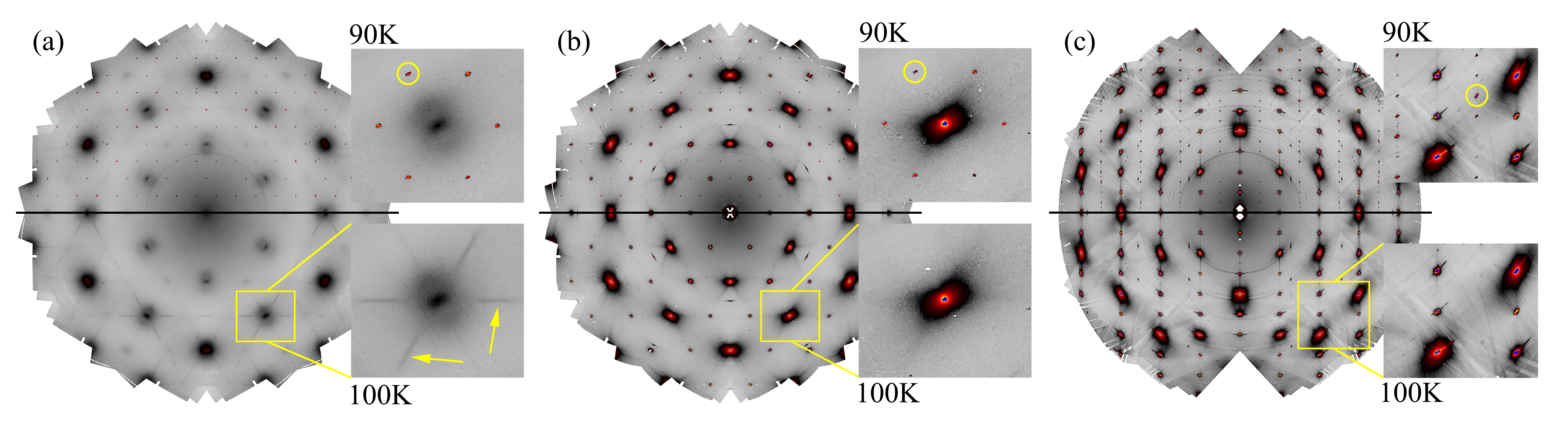}
\caption{\textbf{Diffuse scattering maps of CsV$_3$Sb$_5$}. (a) (\textit{h k} 0.5) plane at 90 K (top), showing the CDW reflections (weak spots highlighted within the yellow circle in the zoomed in area) and the precursor of the 3D CDW at 100 K (bottom), marked with arrows. (b) and (c) (\textit{h k} 0) and (\textit{h} 0 \textit{l}) planes, respectively (no precursor). The DS maps of RbV$_3$Sb$_5$ are shown in the Supplementary Figure 1.}
\label{Fig2}
\end{center}
\end{figure*}

\section{Results}
\subsection{x-ray diffraction}
Figure \ref{Fig1}(a) and (b) sketches the structure of CsV$_3$Sb$_5$ (\textit{P6/mmm} space group). The unit cell is described as individual sublattices of a kagome network of V coordinated by Sb1 atoms and a hexagonal net of Sb2 above and below each kagome layer. The lattice parameters obtained from the DS at room temperature (\textit{a}=\textit{b}=5.49 \AA, \textit{c}=9.34 \AA) are consistent with the reports in the literature \cite{Ort19,Ort20}. The temperature dependence of the CDW instability at the \textit{L} point is displayed in the figure \ref{Fig1}(e), showing a sharp onset at $\sim$95 K, indicative of a first order-like phase transition. Bragg reflections with the same temperature dependence (not shown) were also measured with propagation vectors (0.5 0 0) and (0.5 0 0.25) r.l.u. \cite{Li22Miao}.

\subsection{Diffuse Scattering, DS}
\begin{figure}[h!]
\begin{center}
\includegraphics[width=\columnwidth,draft=false]{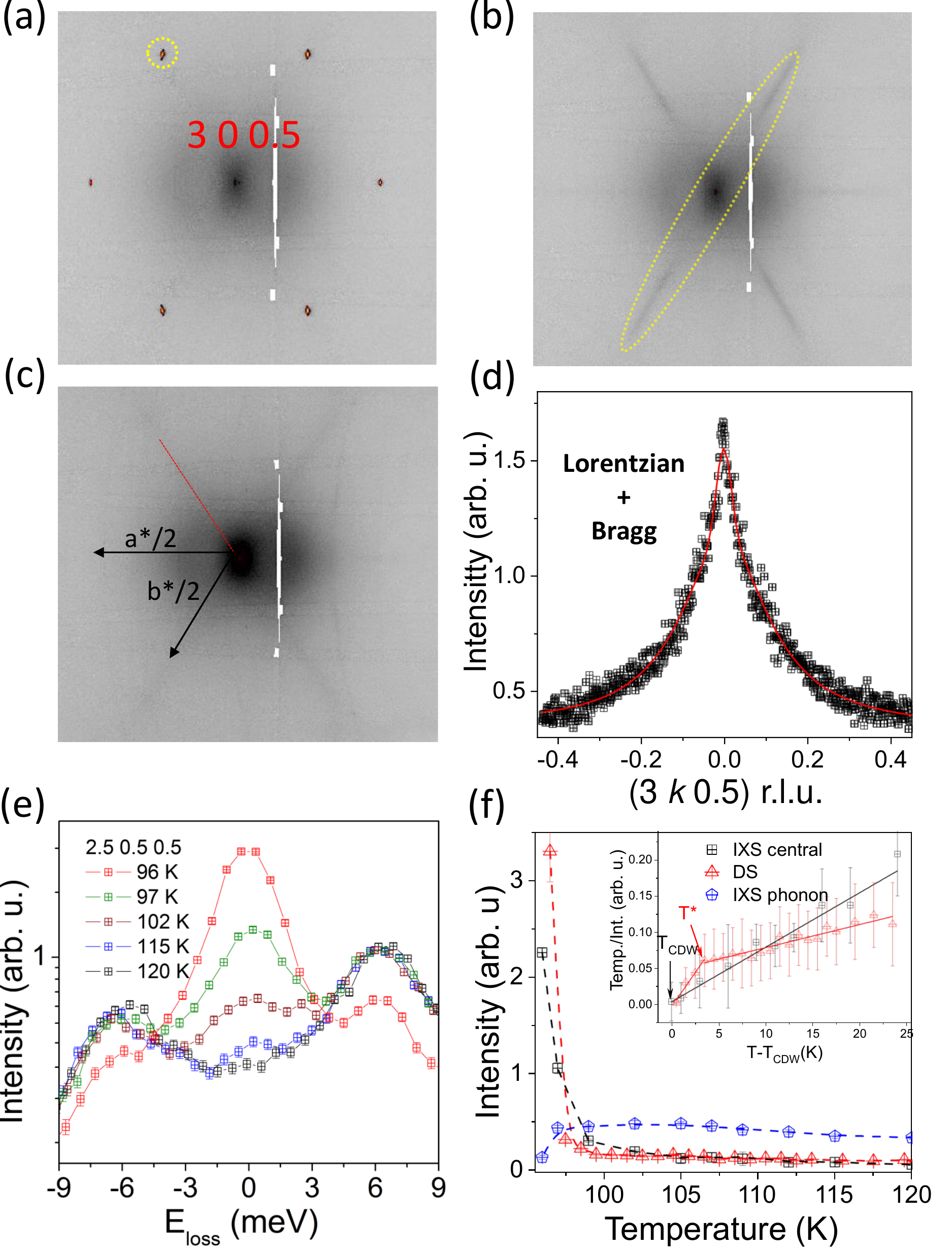}
\caption{\textbf{Temperature dependence of the diffuse scattering and IXS central peak} (a) 90 K, (b) 100 K and (c) 150 K around the (3 0 0.5) cut. The sketch in (c) shows the reciprocal lattice vectors and the IXS scan direction through the CDW, as indicated in the red line. (d) Profile of the diffuse scattering at 120 K and its fitting to a sum of Bragg and Lorentzian profiles. (e) Temperature dependence of the IXS spectra at the (2.5 0.5 0.5) reciprocal lattice vector, highlighting the enhancement of the elastic line and drop of intensity of the low-energy phonon at 96 K. (f) Temperature dependence of the intensity of the DS signal, elastic central peak and optical phonon. Inset, temperature dependence of the generalized susceptibility. T$^*$ denotes the change of slope of the electronic susceptibility, $\sim 105$ K, presumably related to percolation of CDW domains. The error bars are the fit to uncertainty.}
\label{Fig3}
\end{center}
\end{figure}

Besides the sharp and intense CDW diffraction spots in x-ray diffraction, diffuse intensity is present at T$>$T$_{CDW}$ due to lattice vibrations and correlated disorder. Figure \ref{Fig2} summarizes the processed DS maps for the CsV$_3$Sb$_5$ sample at 90 and 100 K for the (\textit{h k} 0.5), (\textit{h k} 0) and (\textit{h} 0 \textit{l}) planes. The image reconstruction of the (\textit{h k} 0.5) plane at 100 K (figure \ref{Fig2}(a) bottom) reveals the presence of DS corresponding to freezing of a transverse component of phonons (streaks highlighted in the yellow square), with most pronounced intensity around the (3 0 0.5) reciprocal lattice vector. Other satellites are equivalent according to the symmetry operations of the \textit{P6/mmm} space group. The DS is visible below 150 K and condense into a 3\textit{Q}$_L$ order at T$_{CDW}$ (figure \ref{Fig2}(a) top). The fluctuating precursor observed in figure \ref{Fig2}(a) along the $<$1 0 1$>$ direction is in agreement with the phonon calculations that predict a lattice instability at the \textit{L} point, figure \ref{Fig1}(c) \cite{Tan21,Subedi22}. The correlation length of the charge precursor ($\xi$) perpendicular to the streak of diffuse intensity remains nearly constant as we cool down to T$_{CDW}$ and extends to 4-5 unit cells, while along the streak reduces to less than one unit cell, \ref{Fig3}(b).  
Remarkably, DS is absent for T$>$T$_{CDW}$ in the (\textit{h k} 0) plane (figure \ref{Fig2}(b) bottom) and (\textit{h} 0 \textit{l}) plane (figure \ref{Fig2} (c) bottom) that maps the longitudinal component of the (0.5 0 0.25) and (0.5 0 0.5) CDWs \cite{Li22Miao,Li22,Jian21}. However, CDW reflections are again present at T$<$T$_{CDW}$, matching the hard x-ray diffraction data of Fig. \ref{Fig1} (e). The presence of diffuse intensity as satellites at only the (0.5 0 0.5) reciprocal lattice vector indicates that the Fermi surface instability at the \textit{L} position is the leading order parameter and drives the CDW formation in the kagome metal AV$_3$Sb$_5$ (see Supplementary Figure 1 for the DS maps of RbV$_3$Sb$_5$).

The reconstruction of the (\textit{h} \textit{k} 0.5) plane for CsV$_3$Sb$_5$ at different temperatures around the (3 0 0.5) cut is shown in figure \ref{Fig3}(a-c). Figure \ref{Fig3}(d) shows the profile of the diffuse precursor along the diffuse streak of the figure \ref{Fig3}(b). The critical scattering was modelled assuming a sum of Bragg and a Lorentzian profile that accounts for the peak sharpening as T approaches T$_{CDW}$ from the high temperature.
The temperature dependence of the DS intensity is extracted by integrating the diffuse precursor and plotted in figure \ref{Fig3}(f). It starts at temperatures below 150 K and increases down to 100-105 K, below which it sharply diverges at the critical temperature, following the temperature dependence of the energy integrated experiment of figure \ref{Fig1}(e).

\subsection{Inelastic x-ray scattering, IXS}
To disentangle static and dynamic contributions and to get further insights about the origin of the observed diffuse features, we have carried out IXS scanning the (2.5 0.5 0.5) satellite. As shown in figure \ref{Fig3}(e), the IXS spectra consists on a (quasi)-elastic central peak (zero energy loss) evolving in intensity below 120 K and a low energy non-dispersive phonon at 6 meV. The intensity of the CP sharply drops upon warming from 95 K and follows the intensity of the pretranslational fluctuations detected by DS above 105 K. The CP of CsV$_3$Sb$_5$ shows parallels to the critical scattering observed in SrTiO$_3$ \cite{Ris71,Sha74} and transition metal dichalcogenides \cite{Mon77} exhibiting a critical divergence on approaching T$_{CDW}$. On the other hand, directly visualized in the raw data of figure \ref{Fig3}(e), the intensity of the low-energy mode increases smoothly upon cooling and drops below 100 K due to the modification of the dynamic structure factor on approaching the CDW phase, making the quasi-elastic contribution more important than the phonon intensity to the DS. Following the applicability of the Ornstein-Zernike correlation function \cite{Rou86}, the generalized susceptibility, $\chi$(q), proportional to the \textit{q} Fourier component of the displacement-displacement correlation function $\langle\lvert$\textit{u$_q$}$\lvert^2\rangle$=\textit{k$_B$T$\chi(q)$}, displays a linear Curie-Weiss behavior for the quasi-elastic component of the CP (inset of figure \ref{Fig3}(f)) and vanishes at the critical temperature, indicating a temperature dependence of the density of states. Nevertheless, $\chi(q)$ of the DS changes slope at T*$\sim$ 105 K, coinciding with the drop of intensity of the low-energy phonon and the deviation of the DS and CP intensities \bibnote{Note1}.      

\begin{figure}[h!]
\begin{center}
\includegraphics[width=\columnwidth,draft=false]{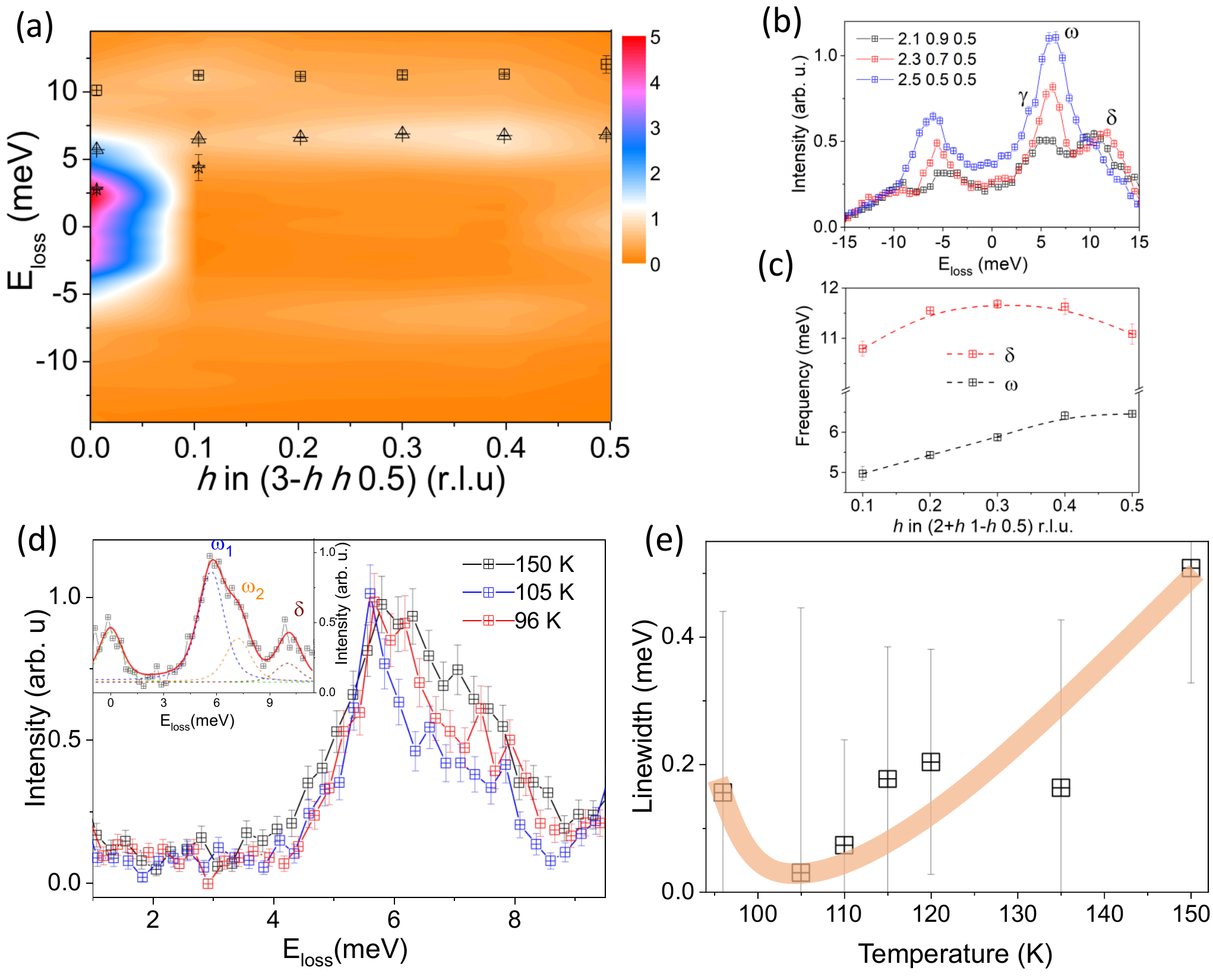}
\caption{\textbf{Einstein modes and phonon broadening} (a) Color map of the low-energy phonons of CsV$_3$Sb$_5$ along the (3-\textit{h h} 0.5) direction at 120 K. The three modes identified at 2.5 ($\gamma$, asterisks), 6 ($\omega$, triangles) and 11 meV ($\delta$, squares) are superimposed. (b) IXS spectra and (c) dispersion along the (2+\textit{h} \textit{1-\textit{h}} 0.5) direction ($\Delta$E= 3 meV) of the $\delta$ and $\omega$ branches. (d) Temperature dependent high resolution IXS spectra ($\Delta$E= 1.5 meV) at the 3.5 2.5 0.5 reciprocal lattice vector. Inset, fitting details of the scan at 150 K. Note that, $\omega$ in (b) is now resolved as 2 modes, $\omega_1$ and $\omega_2$. (e) Temperature dependence of the linewidth of $\omega_1$. The error bars represent the fit uncertainty. See Supplementary for the fitting details and Supplementary Figure 3.}
\label{Fig4}
\end{center}
\end{figure}

Focusing on the inelastic part of the IXS spectra, figure \ref{Fig4}(a) shows the momentum dependence of the phonon dispersion along the (\textit{h h} 0.5) direction ($\Delta$E=3 meV). The spectra consist on 3 weakly dispersive branches with 2.5 ($\gamma$), 6 ($\omega$) and 11 meV ($\delta$) at \textit{h}=0. The lowest energy mode merges with the 6 meV branch at 0.1 r.l.u. and, at the \textit{L} point, only two optical-like modes, $\omega$ and $\delta$, are observed. These two modes are better visualized in figure \ref{Fig4}(b) and its energy dispersion, obtained from the fitting to damped harmonic oscillators (DHO) convoluted with the experimental resolution, is displayed in the figure \ref{Fig4}(c). They exhibit a weak dispersion in reciprocal space and can be described as a system of independent quantum oscillators or localized atomic vibrations (Einstein-type phonon modes). To isolate the contribution of each mode, we increased the energy resolution down to 1.5 meV and work at the (3.5 2.5 0.5) reciprocal lattice vector. Figure \ref{Fig4}(d) shows the temperature dependence of the $\omega$ branch. Indeed, the $\omega$ phonon in figure \ref{Fig4}(b) consists on 2 modes, which we label here as $\omega_1$ and $\omega_2$, in good agreement the DFPT calculations in figure \ref{Fig1}(c) (Supplementary Figure 2). From these data, the energy of $\omega_1$ and $\omega_2$ remains constant as the temperature approaches the CDW transition. This clearly demonstrates the absence of a Kohn anomaly and discards a soft phonon nature of the CDW phase transition; i.e, it is not phonon driven but of order-disorder type. This does not preclude the presence of a soft mode below T$_{CDW}$, which is, in fact, a fingerprint of both order-disorder and displacive phase transitions. We need to stress here that it is not the crystal quality, but the inherent dynamical disorder that stabilizes the CDW in CsV$_3$Sb$_5$. On the other hand, a detailed analysis of the phonon linewidth reveals an anomalous broadening at T$_{CDW}$. At 150 K, $\omega_1$ is not resolution limited and develops a small intrinsic broadening of $\sim$0.5 meV (a DHO with 0 meV intrinsic width corresponds to the instrumental resolution of 1.5 meV). Upon cooling, the width of the phonon decreases and becomes resolution limited at 105 K. However, its width increases again (not resolution limited) below 100 K, coinciding with T*, a signature of a proximity to a Fermi surface instability.    

\section{Discussion}
The results presented here indicate that weak coupling limit theories, $\omega_{2k_F}\tau_{eh}<$1 ($\omega_{2k_F}$ and $\tau_{eh}$ are the bare phonon frequency and electron-hole pair lifetime, respectively), are not adequate to interpret the absence of the Kohn anomaly and the major contribution of the CP to diffuse scattering data. This locates the CDW instability of AV$_3$Sb$_5$ in the strong coupling non-adiabatic or in the strong electron-phonon interaction (EPI) regimes. Indeed, recent reflectivity measurements have reported a much larger CDW gap than the weak-coupling BCS value \cite{Zhou21,Uyk21}.
The effect of the dynamical disorder was theoretically modelled by Yu and Anderson \cite{Yu83} for the strong interaction of electrons and single dispersionless (Einstein-type) optical modes and further extended by Gor'kov \cite{Gor12} for any arbitrary ionic displacements. Within this scenario, the CDW in AV$_3$Sb$_5$ realizes in two stages. First, an ionic displacement in the kagome plane introduces a strong EPI that \textit{traps} the electronic cloud near an ion, changing the local electronic environment and breaking the adiabatic approximation. This gives rise to a quasi-elastic response above the CDW transition, an associated order-disorder dynamics and the absence of a Kohn anomaly. In an order-disorder transformation, the harmonic potential well deforms and the phase transition is achieved by the critical growth and percolation of quasi-static domains on approaching the critical temperature \cite{Cow80}. In each domain, the potential within the atom moves is quasi-harmonic and phonon anomalies are expected. This naturally explains the change of slope in the \textit{diffuse} susceptibility of figure \ref{Fig3}(f) and the small broadening of the low-energy mode in figure \ref{Fig4}(d) and (e). Strong EPI mechanisms have been considered to explain the CDW formation in the one dimensional Peierls instability of BaVS$_3$ \cite{Ila21,Gir19}, transition metal dichalcogenides \cite{Bar75}, H-bond ferroelectrics \cite{Bli06} and the controversial charge order in high-T$_c$ cuprates \cite{LeTacon14}. 
On the other hand, in the non-adiabatic dynamics, $\omega_{2k_F}$ fluctuates so rapidly that cannot couple with the electron–hole condensate during the electron–hole pair lifetime. The rapidly fluctuating phonon reduces the screening between atoms and prevents the development of the Kohn anomaly at \textit{T}$_{CDW}$. Following this argument, the suppression of the Peierls transition of BaVS$_3$ under pressure was interpreted as a reduction of the $\tau_{eh}$ \cite{Ber12}. This is a particularly appealing scenario since the T$_{CDW}$ of CsV$_3$Sb$_5$ is completely suppressed at low pressures and evolves towards a double superconducting dome \cite{Yu21,Zhu22,Yu22,Wang21}.

We would like to finish our discussion by placing our results within the physics of the kagome metals AV$_3$Sb$_5$. The identification of the charge order fluctuations at T$>$T$_{CDW}$ indicates that the condensation of the CDW order parameter at the \textit{L} point in (Cs,Rb)V$_3$Sb$_5$ is the leading instability, favoring the alternating Star of David and trihexagonal (\textit{LLL}) pattern \cite{Fer21}. The absence of DS at the \textit{M} point suggests that the strength of the coupling parameter of the \textit{L} and \textit{M} modes in the Landau free energy expansion, $\gamma_{ML}$, is rather small or the phase transition is more strongly first-order \bibnote{Note2}. Our DS results also indicate that the 6-fold symmetry is preserved, discarding single \textit{Q}$_M$/double \textit{Q}$_L$ orders (staggered SoD and staggered TrH). Rather, our data seem to agree with the experiments that support a combination of triple \textit{Q}$_M$/triple \textit{Q}$_L$ order (\textit{LLL}+\textit{MMM} pattern) \cite{ComNatMat,Luo22a} at low temperature.

On the other hand, the importance of the EPI has been highlighted recently by neutron scattering \cite{Xie22} and, perhaps, acts concomitantly with nesting to enhance the charge correlations. Coherent phonon spectroscopy reported a condensation of \textit{M} and \textit{L}-point phonons, suggesting an order-disorder crossover at lower temperature \cite{Rat21} to a different ordering pattern as observed here. An order-disorder transition is expected to be of first order type, thus the small release of entropy at T$_{CDW}$ \cite{Ort20,Ort21} fits within this scenario. Moreover, the data presented here does not conflict with the possible chiral charge density wave, but put constrains on the nematic order \cite{Nie22,Li22}. Broken rotational symmetries can be affected by local dynamic disorder, fluctuating domains and defects that introduce pinning potentials. These pinning centers and local distortions introduce variations of the amplitudes of the local atomic displacements, which can be detected by pair distribution function (PDF) analysis \cite{Upr22}. 

In conclusion, we have presented diffuse scattering and inelastic x-ray scattering data to demonstrate that the CDW in the kagome metal AV$_3$Sb$_5$ is of order-disorder type without phonon softening. A precursor of the CDW is observed in diffuse scattering with propagation vector (0.5 0 0.5), indicating that the CDW instability at the \textit{L} point is the primary order parameter, discarding a broken six-fold rotational symmetry at high temperature and pointing to an \textit{LLL} pattern. Furthermore, we demonstrate that the diffuse signal contains intensity of the quasi-elastic central peak, likely due to the growth of fluctuating domains, thus favouring strong coupling theories to describe the CDW ground state of AV$_3$Sb$_5$. Our results present a step forward and provide crucial information to understand the enigmatic charge order of (Cs,Rb)V$_3$Sb$_5$.

\section{Methods}
Single crystals of AV$_3$Sb$_5$ (A=Cs, Rb) with T$_{CDW}$=95 and 104 K, were synthesized by the self-flux growth method \cite{Ort19}. Diffuse scattering measurements were performed with a Dectris PILATUS3 1M X area detector at the ID28 beamline at the European Synchrotron Research Facility (ESRF) with an incoming energy \textit{E$_i$}=17.8 keV. Orientation matrix refinement and reciprocal space reconstructions were done using the CrysAlis software package. Final DS images were processed with an in-house code. Low energy phonons were measured by inelastic x-ray scattering (IXS) at the ID28 IXS station at ESRF (\textit{E$_i$}=17.8 keV, $\Delta$E=3 meV) and at the HERIX beamline at the Argonne Photon Source (APS) (\textit{E$_i$}=23.72 keV, $\Delta$E=1.5 meV). The components ($h$ $k$ $l$) of the scattering vector are expressed in reciprocal lattice units (r.l.u.), ($h$ $k$ $l$)= $h \mathbf{a}^*+k \mathbf{b}^*+l \mathbf{c}^*$, where $\mathbf{a}^*$, $\mathbf{b}^*$, and $\mathbf{c}^*$ are the reciprocal lattice vectors. 

Lattice dynamic calculations were performed by Density Functional Perturbation Theory (DFPT) as implemented in QUANTUM ESPRESSO package. The forces were calculated in 8$\times$8$\times$4 supercells using the ultrasoft parametrization of the PBE exchange-correlation functional. We set the energy cutoffs of 60 and 600 Ry for the basis-set and treated the Van der Waals corrections within the Grimme’s semiempirical approach. An 8$\times$8$\times$4 k-point grid and a Marzari-Vanderbilt smearing of 0.01 Ry was used for the Brillouin zone integration. The convergence threshold was set to 10$^{-19}$.

\section{Data availability}
Source data are provided with this paper. The DS and IXS data generated in this study have been deposited in the Figshare database https://doi.org/10.6084/m9.figshare.21990461.

\section{Acknowledgments}
We acknowledge E. Efremov, A. Schnyder, A. Bernevig, A. Subedi, J. Diego, L. Elcoro, I. Etxebarria, D. Chernyshov and R. Fernandes for fruitful discussions and critical reading of the manuscript. S.B-C, D.S and L.S thank the MINECO of Spain through the project PGC2018-101334-A-C22. S.D.W and B.R.O gratefully acknowledge support via UC Santa Barbara NSF Quantum Foundry funded via the Q-AMASE-i program under award DMR-1906325. This research used resources of the Advanced Photon Source, a U.S. Department of Energy (DOE) Office of Science user facility operated for the DOE Office of Science by Argonne National Laboratory under Contract No. DE-AC02-06CH11357.

\section{Author Contribution}
S.B-C conceived and managed the project. B.O and S.D.W. synthesized and characterized the samples. A.K., A.B. and S.B-C performed the DS experiments. D.S., A.K., A.H.S., L.S., A.B. and S.B-C carried out the IXS measurements. S.B-C performed the DFPT calculations, analyzed the data and wrote the manuscript with input from all coauthors.

\section{Competing Interests}
The authors declare no competing interests.
\end{document}